\documentclass[review]{elsarticle}

\def\KAOS{{\sc Kaos}\@}
\usepackage{graphicx}
\usepackage{amssymb}
\usepackage{lineno}

\journal{Nucl. Instr. and Meth. in Phys. Res. A}

\begin{document}

\begin{frontmatter}

\title{Characterisation and calibration of a scintillating fibre
  detector with $>$ 4\,000 multi-anode photomultiplier channels}

\author{A.~Esser\corref{cor}} \ead{aesser@kph.uni-mainz.de}
\author{P.~Achenbach\corref{cor}} \ead{patrick@kph.uni-mainz.de}
\author{C.~{Ayerbe Gayoso}} \author{M.~Biroth} \author{P.~G\"ulker}
\author{J.~Pochodzalla}

\address{Institut f\"ur Kernphysik, Johannes Gutenberg-Universit\"at
  Mainz, Germany}

\cortext[cor]{Corresponding authors. Tel.: +49-6131-3925831
  (-3925826); fax: +49-6131-3922964.}

\begin{abstract}
In the \KAOS\ spectrometer at the Mainz Microtron a high-resolution
coordinate detector for high-energy particles is operated. It consists
of scintillating fibres with diameters of $<$ 1\,mm and is read out by
$>$ 4\,000 multi-anode photomultiplier channels.  It is one of the
most modern focal-plane detectors for magnetic spectrometers
world-wide.

To correct variations in the detection efficiency, caused by the
different gains and the different optical transmittances, a fully
automated off-line calibration procedure has been developed.  The
process includes the positioning of a radioisotope source alongside
the detector plane and the automated acquisition and analysis of the
detector signals. It was possible to characterise and calibrate each
individual fibre channel with a low degree of human interaction.
\end{abstract}


\end{frontmatter}


\section{The scintillating fibre detector}

\begin{figure}
  \centering
  \includegraphics[width=0.9\columnwidth]{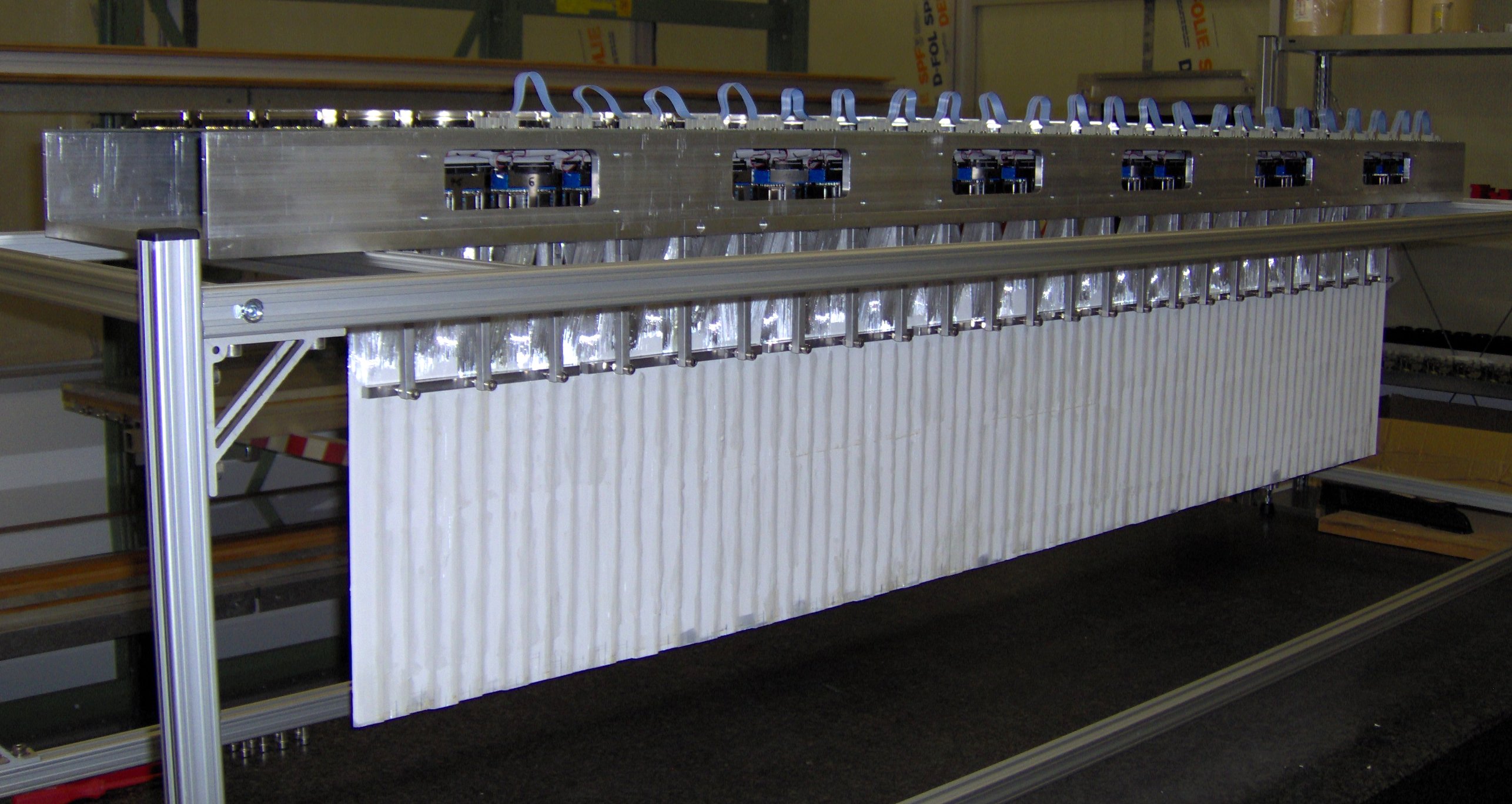}
  \caption{Photograph of one detector plane with 9\,216 fibres, 72
    multi-anode PMTs and 2\,304 read-out channels. Each individual
    channel was characterised and calibrated. The coordinate detector
    consists of two such planes.}
  \label{fig:photo_plane} 
\end{figure}

The high resolution coordinate detector in the \KAOS\ spectrometer
operated at Mainz Microtron consists of two identical planes of 9\,216
Kuraray SCSF-78 scintillating fibres of 0.83\,mm outer diameter.  Each
plane is read out by 72 Hamamatsu H7259K multi-anode photomultipliers
(MaPMT) with 32 channels in a linear array.  The MaPMT have an average
anode luminous sensitivity of $S_a= 374\,$A$/$lm (according to data
sheet 140\,A$/$lm is typical), an average cathode luminous sensitivity
$S_k=84\,\mu$A$/$lm (70\,$\mu$A$/$lm is typical), and an average gain
$G= 4.4 \cdot 10^6$ ($2\cdot 10^6$ is typical).
Fig.~\ref{fig:photo_plane} shows a photograph of one detector plane
during the assembly stage.

Individual Cockcroft-Walton voltage multipliers provide the
high-voltages for the MaPMTs~\cite{dubna:hvsys}.  The high voltage is
distributed linearly between the 10 stages of metal channel dynodes.
The signals of the 144 MaPMTs are transmitted by 96-channel front-end
boards and digitised by double-threshold discriminators.  The position
information and signal time are picked up by state-of-the-art ${\cal
  F}$1 time-to-digital
converters~\cite{Achenbach2009:FrontEndElectronics}.  A fast trigger
logic based on a total of 37 FPGA modules controls the data
acquisition.  This DAQ system is nearly dead-time
free~\cite{Achenbach2010:RT10}.

This detector combines a precise charged particle track determination
with a high resolution timing information and can tolerate background
particle rates of the order of several hundred MHz.

\section{Calibration setup}

\begin{figure}
  \centering
  \includegraphics[width=0.9\columnwidth]{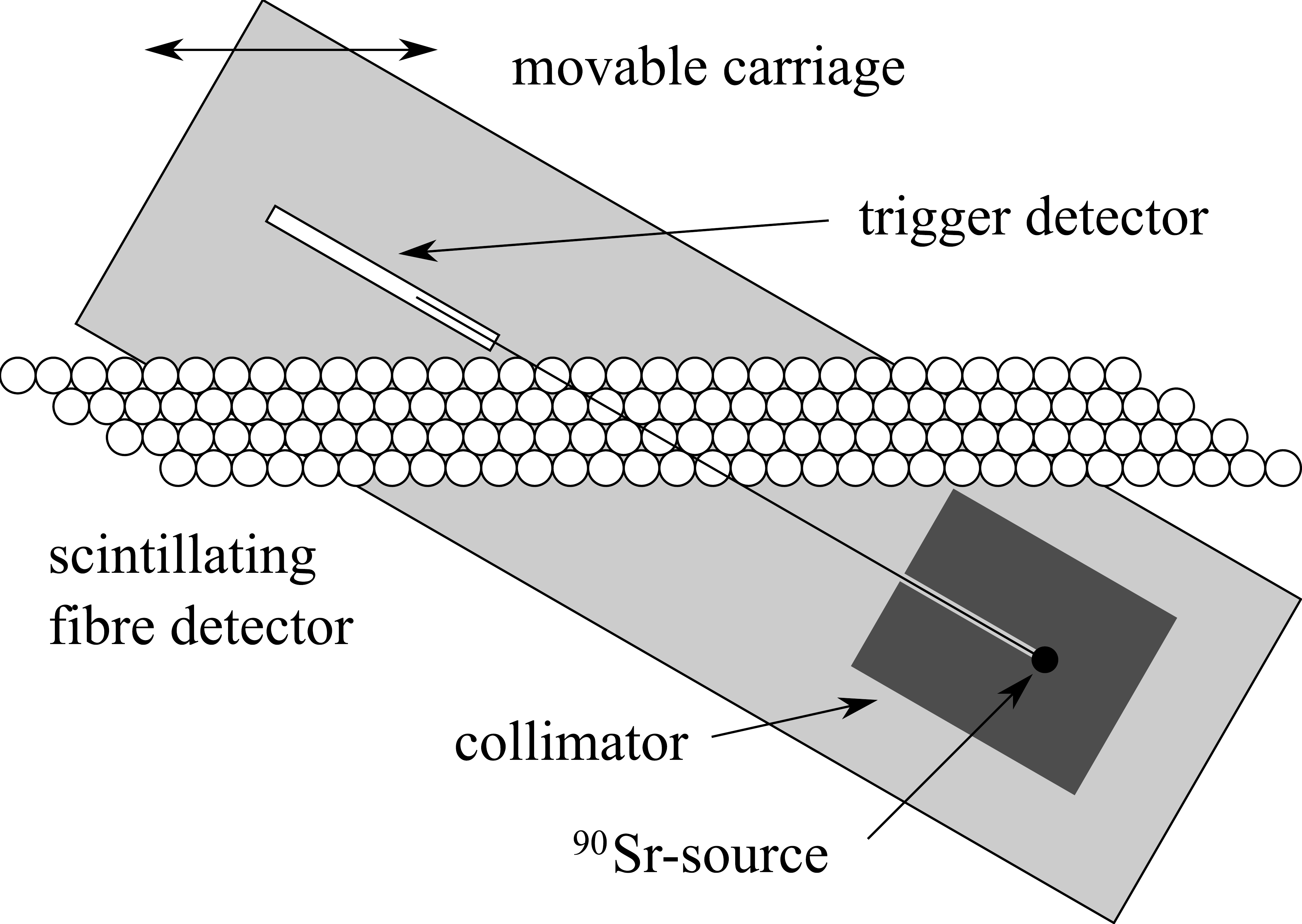}
  \caption{Schematic drawing of the calibration setup. A collimated
    $\beta$-source and a trigger detector are mounted on common
    carriage for automatic positioning alongside the coordinate
    detector plane.}
  \label{fig:setup} 
\end{figure}

To correct variations in the detection efficiency, caused by the
different gains of the MaPMTs and variations of the optical
transmittance and coupling of the fibres to
the photocathode, a fully automated calibration procedure has been
developed.

A collimated $^{90}$Sr-source with a beam diameter of 1\,mm and a
total activity of 22 MBq is placed in an acrylic glass block.  On the
opposite side a narrow trigger detector consisting of a stack of
scintillating fibres assures that the electrons have crossed the
fibre detector.  This trigger detector has a size of L $\times$ W
$\times$ H $=$ 20 $\times$ 0.5 $\times$ 20 mm$^3$ and is read out by a
conventional PMT.  The setup is mounted on a linear guidance and is
positioned alongside the detector plane by a stepping motor with a
precision of 0.1\,mm.  A scheme of the calibration setup can be seen in
Fig.~\ref{fig:setup}.

To normalise the energy loss of the electrons inside the fibre
detector, the trigger detector discriminator threshold is chosen so
high that only the most energetic particles are accepted thus requiring
minimum ionisation in the fibre detector.

\section{Characterisation and alignment}
\begin{figure}
  \centering
  \includegraphics[width=0.9\columnwidth]{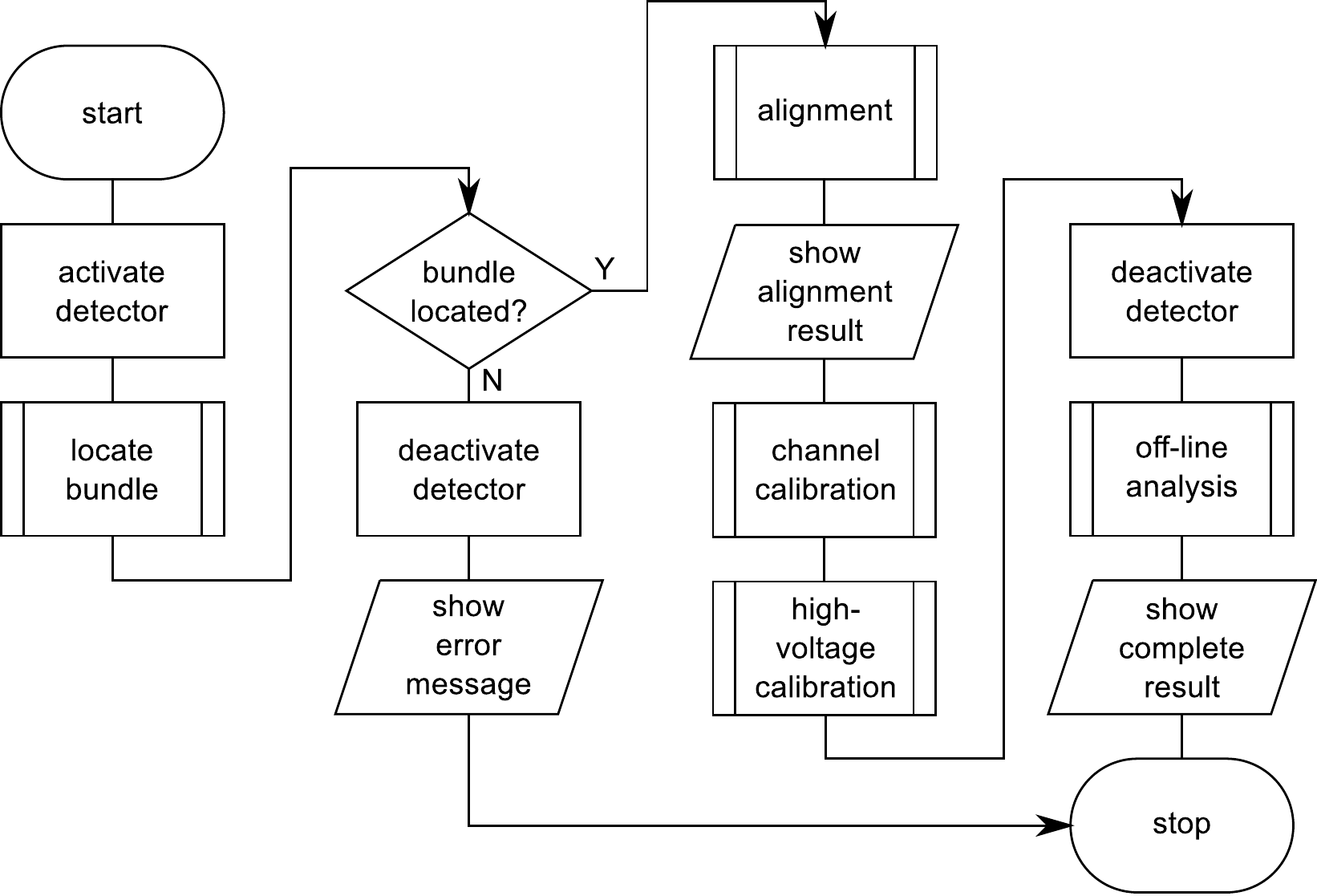}
  \caption{Process flow of the calibration procedure.  After an
    alignment of the calibration setup to the fibre detector, a
    channel by channel calibration measurement is conducted and the
    high-voltage response is measured.  All data is stored for an
    off-line analysis.}
  \label{fig:flowchart} 
\end{figure}

The process of calibration and characterisation for a detector section
of 32 channels coupled to one MaPMT is divided into three major steps
as shown in Fig.~\ref{fig:flowchart}.  First, an alignment between the
detector coordinates and the calibration setup is found.  This
provides the position information for the channels in this section.
Second, the gain measurement of all 32 channels is performed.
Finally, the gain variation with the high-voltage for this particular
MaPMT is determined.  All collected data is stored for off-line
analysis.

\subsection{Position alignment}

\begin{figure}
  \centering
  \includegraphics[angle=270,width=0.9\columnwidth]{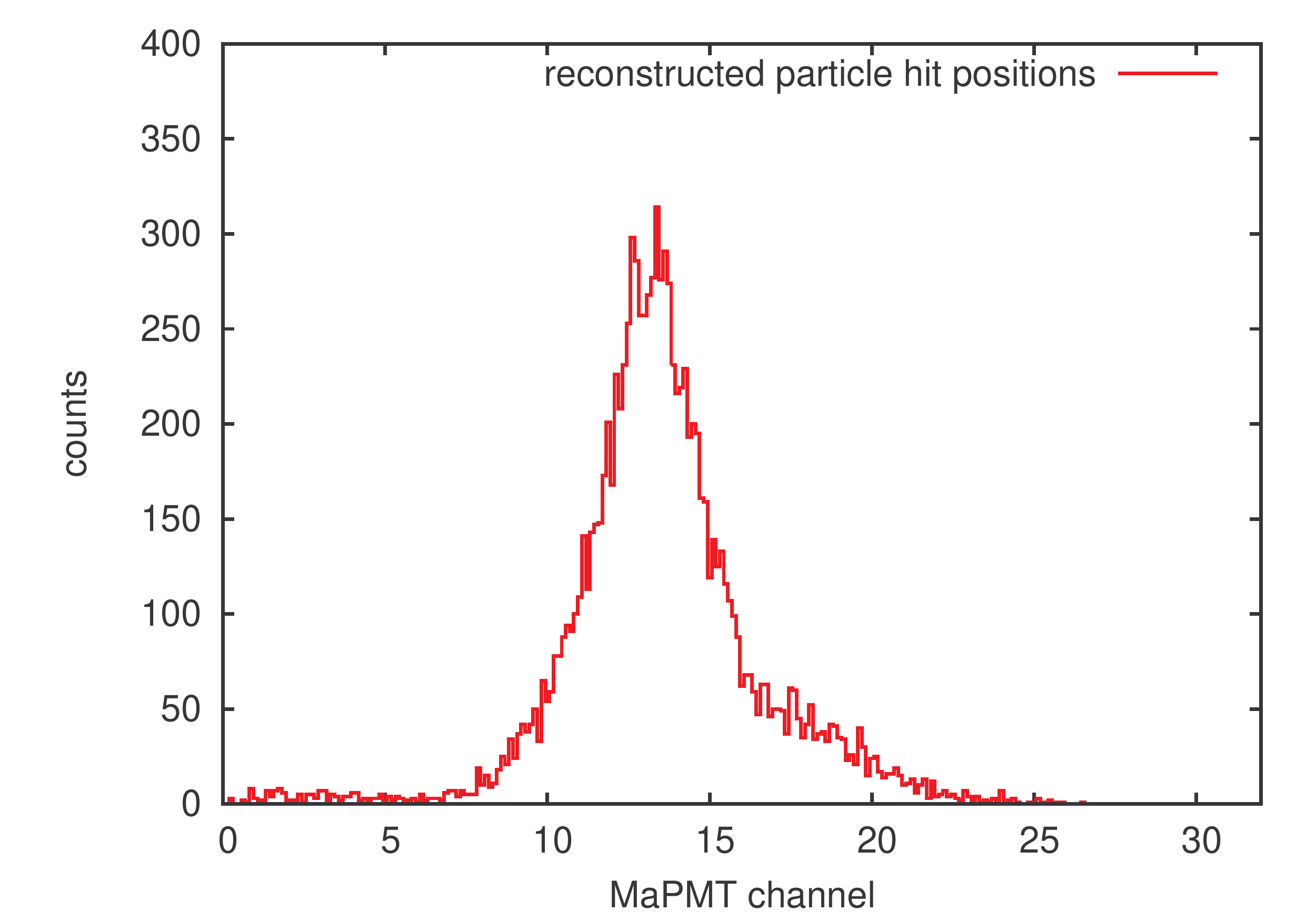}
  \caption{Typical hit position distribution of a measurement with the
    calibration setup positioned at channel 13.  One channel 
    corresponds to 0.83 mm.  The tail of the
    distribution between channels 16 and 22 is caused by the tilted
    channel arrangement causing a higher probability for particles to
    be scattered into the trigger detector from the higher channel
    side.}
  \label{fig:position} 
\end{figure}

\begin{figure}
  \centering
  \includegraphics[angle=270,width=0.9\columnwidth]{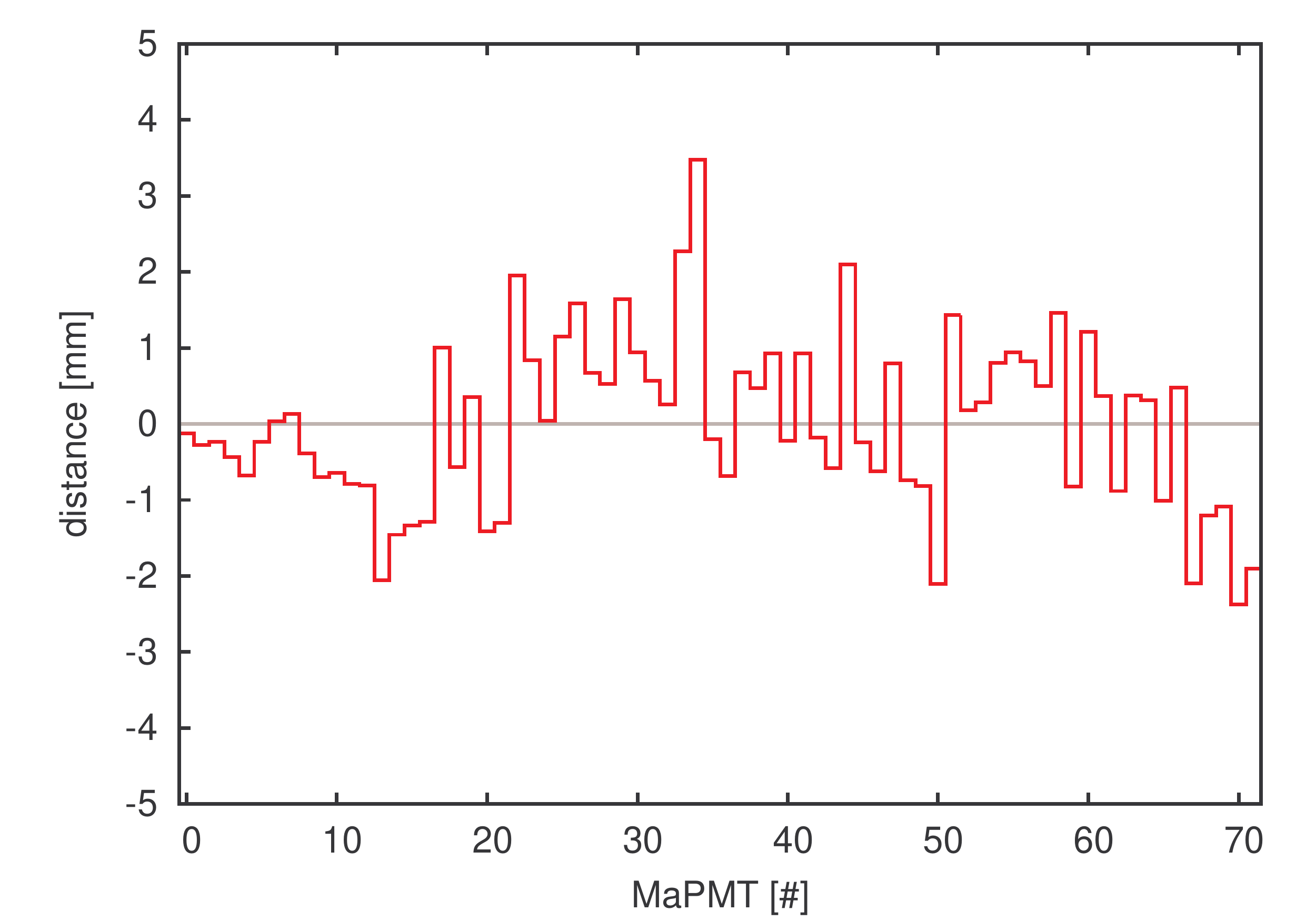}
  \caption{Deviations of the channel positions for all 72
    individual fibre bundles with respect to a linear channel
    arrangement.  This data is used for correcting the channel
    positions in the coordinate detector data analysis.}
  \label{fig:allAlignments} 
\end{figure}

Since the detector will be used for tracking particles, it is
essential to know the precise position of every channel.  Each
detector plane is assembled from bundles of 128 fibres to allow for
easy replacement of parts of the plane.  The fibres of one individual
bundle show very small position variations since they are pressed
together during the building process by a precise aluminium matrix.
However, gaps between the fibres of two neighbouring MaPMTs can be of
the order of 1\,mm.

After the calibration software has located one fibre bundle by moving
the carriage to its approximate position and taking a test
measurement, the first step is to measure its position.
Therefore the source is placed in front of the bundle at 10
equidistant positions between channel 6 and 24.  For every position a
data set of 2\,000 events is acquired.  From this data the hit
positions are reconstructed by analysing signal clusters.  These
clusters are caused by particles hitting scintillating fibres of
multiple channels as well as optical cross-talk between neighbouring
channels.  The hit position is calculated as the mean channel of the
cluster where all channels are weighted with their ADC value.  This
allows to achieve a position resolution below the channel width.  The
position of the radiation source is used as a reference point.

A typical position measurement is shown in Fig.~\ref{fig:position}.
The sizes of the collimated electron beam and the trigger detector
cause the width of this distribution, the asymmetry of the peak 
shape is caused by the tilted channel arrangement of the fibre detector,  
which increases the likelihood of particles on the higher
channel side to be scattered into the trigger detector.  This
systematic error only results in a uniform shift of all channel
positions and does not contribute to the position resolution.

The obtained positions of the individual bundles of scintillating
fibres is used to reconstruct the precise hit position in the analysis
of the data from the coordinate detector in real operation.  The
relative deviations of the fibre bundle positions with respect to a linear
arrangement is shown in Fig.~\ref{fig:allAlignments}.

\subsection{Gain measurement}

\begin{figure}
  \centering
  \includegraphics[angle=270,width=0.9\columnwidth]{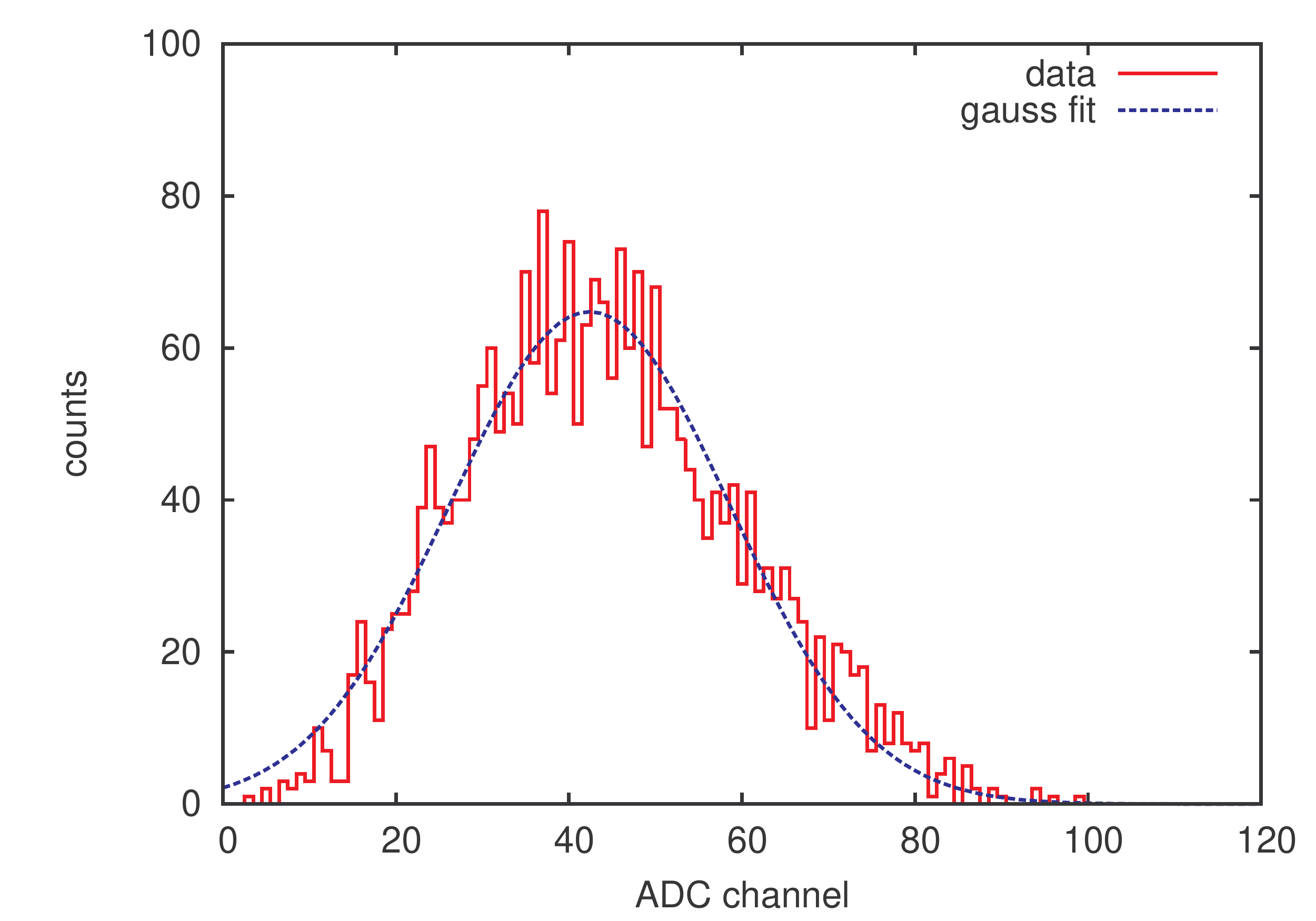}
  \caption{Typical ADC value distribution for the signal height
    measurement of one channel.  The relative energy resolution is
    $\Delta E/E\approx$ 0.6, leading to a measurement accuracy of
    approximately 1\%.}
  \label{fig:adcHisto} 
\end{figure}

\begin{figure}
  \centering
  \includegraphics[angle=270,width=0.9\columnwidth]{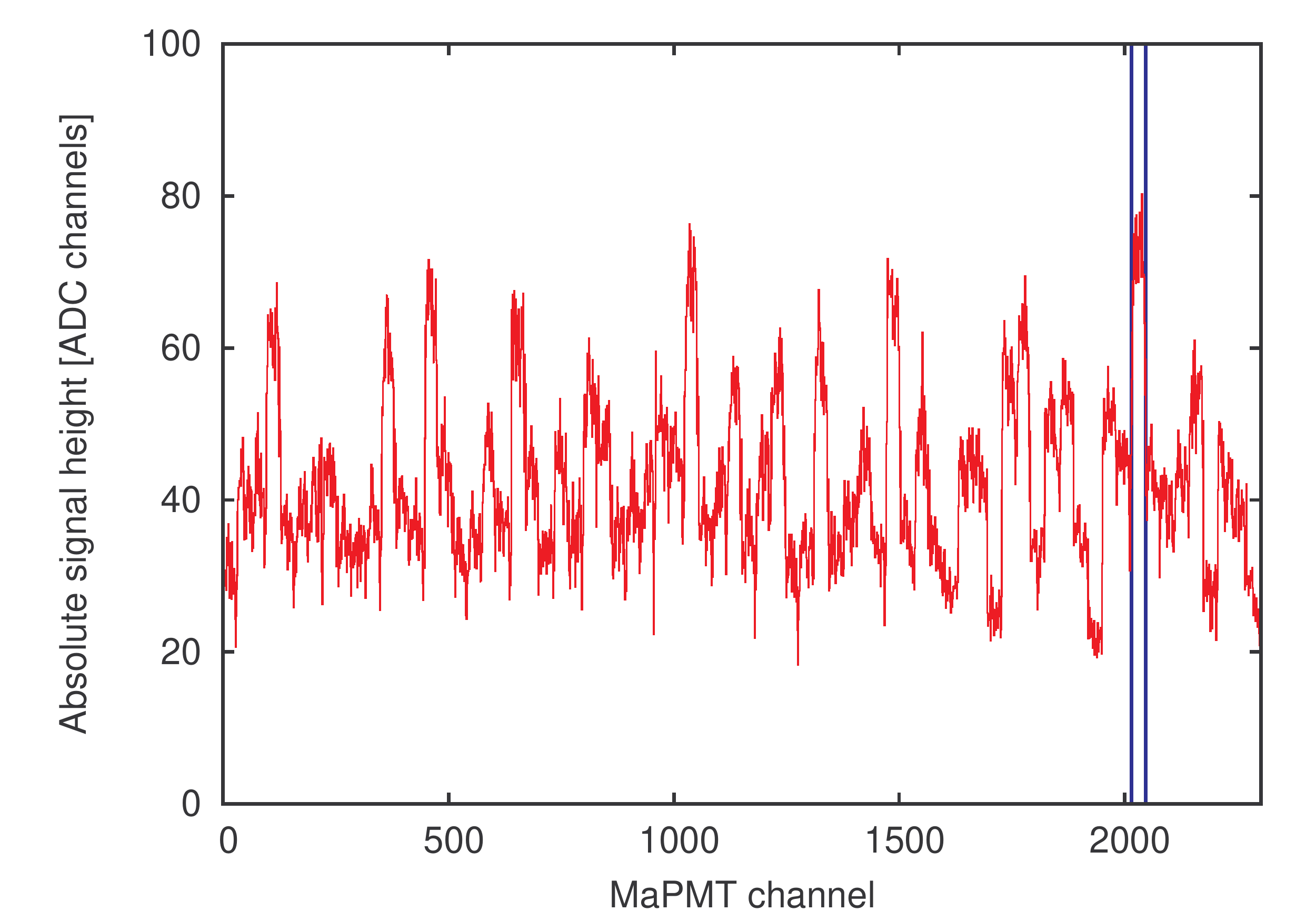}
  \caption{Measured signal heights for all 2\,304 channels of one
    detector plane.  The vertical lines indicate the channels covered
    by one MaPMT.  Large variations between different MaPMTs, as well
    as smaller variations between the channels of one MaPMT can be
    observed.}
  \label{fig:allChannels} 
\end{figure}

After the position of a bundle has been measured, in the next step the
individual gain of every channel is determined.  Therefore the
assembly of trigger detector and source is placed to every channel and
10\,000 events of data are collected at a fixed high-voltage of 900\,V.

To achieve a good energy resolution it has to be assured that the
entries in the ADC spectrum correspond to particles which have crossed
similar track lengths inside the active fibre cores.  Therefore, only
events with hit positions in the range of the regarded channel
contribute to the spectrum.  An example for a resulting ADC spectrum
is shown in Fig.~\ref{fig:adcHisto}. The relative energy resolution
which can be achieved by this method of data acquisition and analysis
is of the order of $\Delta E/E\approx$ 0.6. This leads to an accuracy
for these measurements of the order of 1\%.

In the signal heights for all 2\,304 channels of one plane as displayed
in Fig.~\ref{fig:allChannels}, large gain variations of the order of
40\% can be observed between individual MaPMTs, smaller variations of
the order of 10\% appear between the channels of one MaPMT.  
According to the data sheet, the gain uniformity between
anodes is specified to be between 1:1.1 and 1:1.25 (1:1.5 typical),
with the edge anodes having slightly lower gains on average.
Few channels have less than 70\% of the maximum gain of the MaPMT.
However the gain values specified in the data sheet do not correspond
to the measured values, since the measurement also takes into account
the optical properties of each channel.

\subsection{High-voltage adjustment}

\begin{figure}
  \centering
  \includegraphics[angle=270,width=0.9\columnwidth]{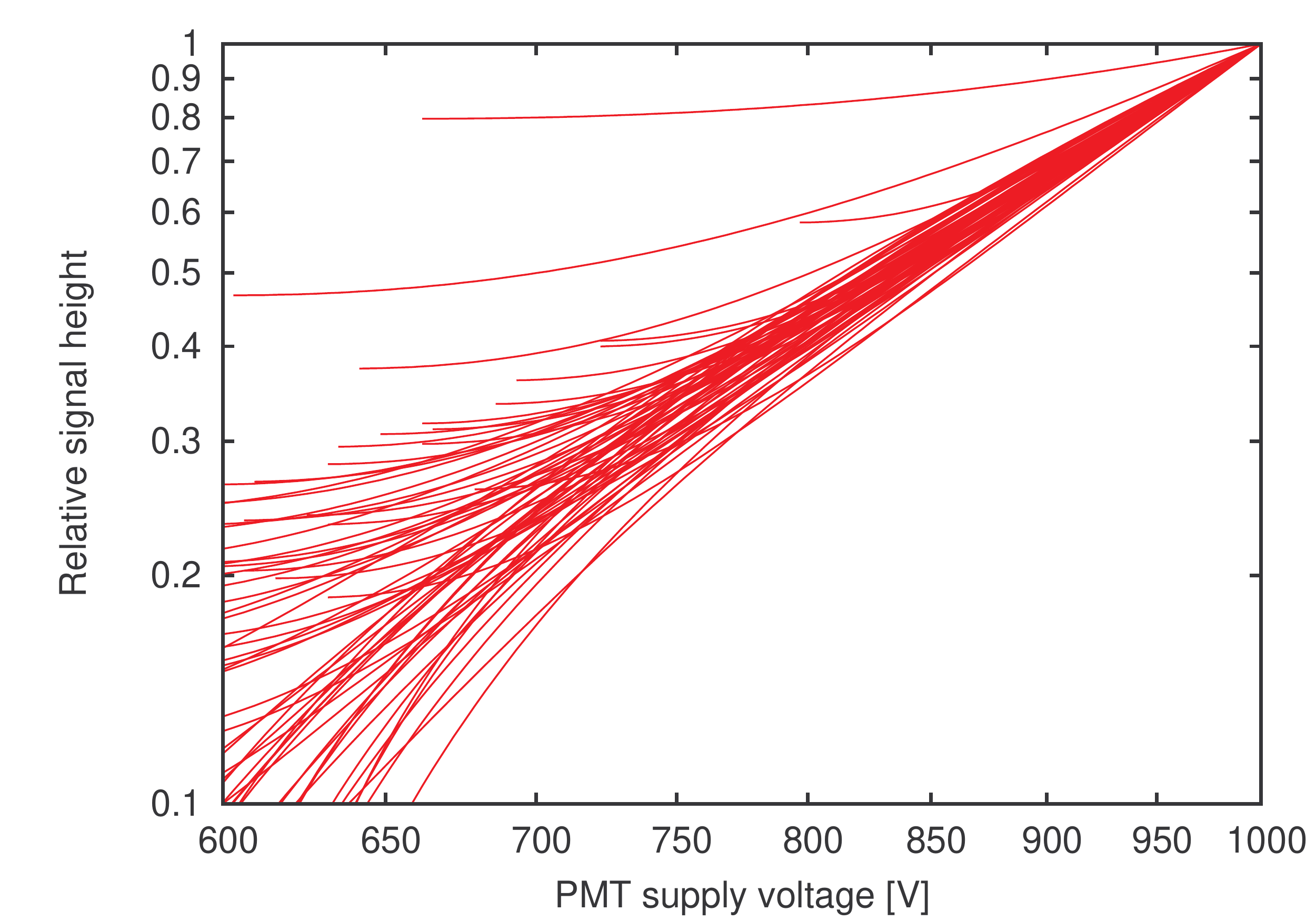}
  \caption{Relative signal height as a function of the applied
    high-voltage for 72 MaPMTs.
    The characteristics of individual MaPMTs can be distinguished from
    the general trend.}
  \label{fig:relativeGain} 
\end{figure}

\begin{figure}
  \centering
  \includegraphics[angle=270,width=0.9\columnwidth]{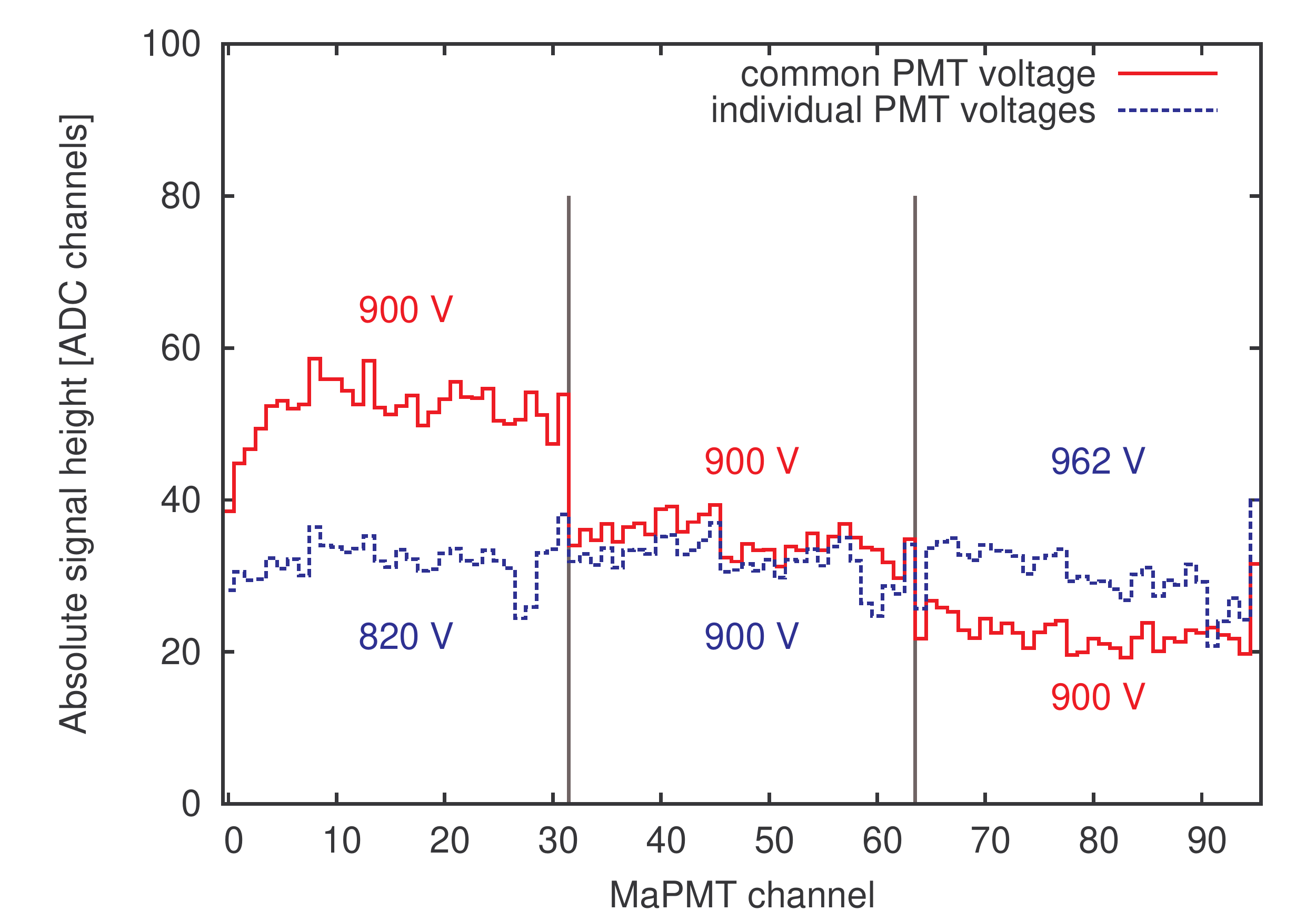}
  \caption{Measured signal heights of the 96 channels from three
    neighbouring MaPMTs when operated with a common high-voltage of
    900\,V (red, solid), and after an individual high-voltage
    adjustment (blue, dashed).  The high-voltages are adjusted so that
    the ADC values of the channels with the highest gain of each MaPMT
    are equal.}
  \label{fig:adjustment} 
\end{figure}

The signal height of the two channels with the highest gain and the
channel with the lowest gain is measured for each MaPMT with 9
different high-voltages in the range of 600\,V to 1000\,V.  The
relative signal height variation with the high-voltage shows a similar
behaviour for all 72 PMTs of one detector plane as shown in
Fig.~\ref{fig:relativeGain}.  The signal height changes proportional
to the 4.5$^{th}$ power of the high-voltage.  The gain increase
specified in the data sheet is proportional to the 9$^{th}$ power of
the high-voltage.  However, the manufacturer measures the gain as the
ratio between the anode current and the current from the photocathode,
while the signals in this calibration procedure are charge integrated
single anode pulses.  Some MaPMTs with significantly lower gain than
the average can be distinguished from the general trend.  For these
MaPMTs the signal height increase with high-voltage is lower.

With the response measured for all MaPMTs the high-voltages can
be adjusted to correct the gain variations between different MaPMTs.
The high-voltages are adjusted so that the ADC values of the channels
with the highest gain of each MaPMT are equal.
The effect of this
adjustment can be seen in Fig.~\ref{fig:adjustment}.  Variations
between different channels of one MaPMT are compensated by setting
individual discriminator thresholds.

\subsection{Identification of faulty channels}

During the first calibration runs several faulty channels were
identified by missing entries in the ADC spectra.  The main reasons
were broken connectors and cables between the MaPMT bases and the
front-end boards which retransmit the signals to the discriminators.
The modular detector construction allowed to replace all faulty parts.

\section{Conclusion}
The high number of channels leading to a total of almost 9\,000
measurements needed to characterise this detector made it necessary to
implement a setup that works highly autonomously.  Using this setup
all data was collected which is necessary to set the high-voltages for
the MaPMTs and discriminator thresholds to achieve equal detection
efficiency in all channels, and to correct the channel positions in
the off-line data analysis to maximise the spatial resolution
of the charged particle track determination.  The
only form of human interaction needed with the setup is the rewiring of the
MaPMTs and the manual repair of broken connectors leading to faulty
channels.

\section*{Acknowledgements}

This work was supported in part by the Federal State of
Rhineland-Palatinate and by the Deutsche Forschungsgemeinschaft with
the Collaborative Research Center 443.

We acknowledge the support by the Research Infrastructure
Integrating Activity ``Study of Stongly Interacting Matter''
HadronPhysics2 under the 7th Framework Programme of EU.

The assistance of staff, particularly from the mechanics and
electronics workshops, is gratefully acknowledged.

Part of this work is contained in the PhD thesis of C. Ayerbe Gayoso.

\bibliographystyle{elsarticle-num}
\bibliography{proceeding}

\end{document}